\documentclass[twocolumn,preprintnumbers,superscriptaddress,nofootinbib,aps,prd,floatfix]{revtex4}

\usepackage{footmisc,multirow,amssymb}
\usepackage{subfigure}
\usepackage{amsmath,slashed,enumerate}
\usepackage{color}
\usepackage{graphicx}

\newcommand{\be}{\begin{eqnarray*}}
\newcommand{\ee}{\end{eqnarray*}}

\newcommand{\bee}{\begin{eqnarray}}
\newcommand{\eee}{\end{eqnarray}}
\newcommand{\beeq}{\begin{equation}}
\newcommand{\eeeq}{\end{equation}}

\preprint{IPPP/16/56} \preprint{DCPT/16/112}

\begin{document}

\title{Towards resolving strongly-interacting dark sectors at colliders}

\begin{abstract}
Dark sectors with strong interactions have received considerable interest. Assuming the existence of a minimally coupled dark sector which runs to strong interactions in the infrared, we address the question whether the scaling behavior of this dark sector can be observed in missing energy signatures at present and future hadron colliders. We compare these findings to the concrete case of self-interacting dark matter and demonstrate that the energy dependence of high-momentum transfer final states can in principle be used to gain information about the UV structure of hidden sectors at future hadron colliders, subject to large improvements in systematic uncertainties, which could complement proof-of-principle lattice investigations. We also comment on the case of dark Abelian $U(1)$ theories.
\end{abstract}

\author{Christoph Englert} \email{christoph.englert@glasgow.ac.uk}
\affiliation{SUPA, School of Physics and Astronomy, University of
  Glasgow,\\Glasgow, G12 8QQ, United Kingdom\\[0.1cm]}
  
\author{Karl Nordstr\"om} \email{k.nordstrom.1@research.gla.ac.uk}
\affiliation{SUPA, School of Physics and Astronomy, University of
  Glasgow,\\Glasgow, G12 8QQ, United Kingdom\\[0.1cm]}

\author{Michael Spannowsky} \email{michael.spannowsky@durham.ac.uk}
\affiliation{Institute for Particle Physics Phenomenology, Department
  of Physics,\\Durham University, Durham, DH1 3LE, United Kingdom\\[0.1cm]}

\maketitle

%%%%%%%%%%%%%%%%%%%%%%%%%%%%%%%%%%%%%%%%%%%%%%%%%%

\section{Introduction}
\label{sec:intro}

Astronomical observations strongly indicate the existence of dark matter~\cite{Komatsu:2010fb,Ade:2013zuv}. Many extensions of the Standard Model (SM) take this into account by incorporating a so-called dark sector: a sector of particles that are not charged under the SM gauge interactions. The interactions of the dark sector can be protected by global symmetries and the particles can have a long lifetime, thus providing plausible dark matter candidates. Phenomenologically nontrivial and hence collider-relevant theories do not completely decouple the dark sector, but introduce interactions with the Standard Model fields through the exchange of a mediator of a yet unknown force. At the renormalisable level such an interaction can be facilitated by $U(1)$ mixing~\cite{Holdom:1985ag,Foot:2014uba,Foot:2016wvj} or a so-called Higgs portal interaction for the SM field content~\cite{Binoth:1996au,Schabinger:2005ei,Patt:2006fw}.

Recently, dark sectors with strongly interacting particles have become of interest. Nonobservation of smoking gun signatures predicted by solutions inspired by the WIMP miracle has spurred dark matter (DM) model building to explore different directions, further supported by astrophysical measurements that could be explained by complex, non-weakly-interacting dark sectors. 

One avenue is to predict the existence of self-interacting dark matter, thereby addressing e.g. the {core vs cusp}~\cite{deBlok:2009sp}, {too-big-to-fail}~\cite{BoylanKolchin:2011de}, {missing satellite}~\cite{Bullock:2010uy} and Tully-Fisher Galaxy Halo~\cite{Tully:1977fu,Moore:1999nt,Burkert:1995yz} problems simultaneously. The existence of complex dark sectors is further motivated by the fact that there is no \textit{a priori} reason why dark matter interactions should exhibit a trivial interaction structure in comparison to the SM, which only makes up for 15\% of the Universe's baryonic matter content. 
In self-interacting DM scenarios an energy transfer from the outer hotter region of the halo to the central colder region can produce a core structure in agreement with current observations. In addition, the number of Milky Way satellite galaxies is significantly reduced. However, to allow self-interaction of dark matter particles to explain these observations without being excluded by others requires a large interaction cross section of $0.1~\mathrm{cm}^2/\mathrm{g} \leq \sigma/m  \leq 10~\mathrm{cm}^2/\mathrm{g}$~\cite{Dave:2000ar,Vogelsberger:2012ku,Rocha:2012jg,Peter:2012jh}, where $m$ denotes the dark matter candidate mass. 

A strongly interacting, potentially nonminimal, dark sector can give rise to composite DM~\cite{Nussinov:1985xr,Barr:1990ca,Khlopov:2005ew,Gudnason:2006ug,Gudnason:2006yj,Khlopov:2008ty,Ryttov:2008xe,Foadi:2008qv,Alves:2009nf,Mardon:2009gw,Kribs:2009fy,Frandsen:2009mi,Lisanti:2009am,Khlopov:2010pq,Belyaev:2010kp,Lewis:2011zb,Buckley:2012ky,Hietanen:2012qd,Hietanen:2012sz,Appelquist:2013ms,Hietanen:2013fya,Cline:2013zca,Boddy:2014yra,Boddy:2014qxa,Cohen:2015toa} and dark atoms~\cite{Alves:2010dd,Behbahani:2010xa,Kaplan:2011yj,Kumar:2011iy,Khlopov:2011tn,Cline:2012is,CyrRacine:2012fz,Fan:2013yva,Fan:2013tia,McCullough:2013jma,Bai:2013xga,Cline:2013pca,Belotsky:2014haa,Hansen:2015yaa,Arthur:2016dir} resulting in a rich phenomenology. 

While the very existence of dark matter is a strong indication of the presence of a secluded sector, dark sectors are also motivated beyond the realm of dark matter. Portal-type interactions have been motivated in generic hidden valley scenarios~\cite{Strassler:2006im,Han:2007ae}, dark energy model building~\cite{Krauss:2013oea,Burrage:2016xzz}, as well as conformal SM extensions to tackle the hierarchy problem~\cite{Meissner:2006zh,Englert:2013gz,Abel:2013mya,Latosinski:2015pba} with potential links to leptogenesis~\cite{Khoze:2013oga} and inflation~\cite{Khoze:2013uia}. Dark sectors and their interaction with the SM spectrum can therefore be considered as versatile tools to tackle apparent shortcomings of the SM, typically leading to the production of new states in high-energy interactions.

Irrespective of their motivation, we are faced with the question of how much we can learn about dark sectors by the very fact that their interaction with visible sectors is suppressed. How much can present and future high-energy colliders contribute to a resolution of this question? 

It is known that inclusive rates of $Z$ and Higgs boson interactions, as well as new resonances in multi-Higgs final states, can be indicative of mixing effects with dark sectors~\cite{Dolan:2012ac,Choi:2013qra,Papaefstathiou:2015paa}. However, depending on the complexity of mediator interactions and dark sectors, it proves very difficult to enable a comprehensive dark sector ``spectroscopy''~\cite{Dreiner:2012xm,Andersen:2013rda,Chacko:2013lna,Becciolini:2014lya,Alves:2014cda,Khoze:2015sra}. In this paper we show that some information about the strong dynamics of a hidden sector can be gained by studying the momentum dependence of telltale $E_T^\text{miss}$ events. Although we limit ourselves to $H+\text{jet}$ final states in the following, our arguments apply to any process that involves mediator production at high-energy colliders.

This work is organised as follows: We first review typical mediator scenarios and discuss in how far strong dynamics can leave visible phenomenological footprints through renormalisation group effects at colliders in Sec.~\ref{sec:model}, before we focus on a minimal scenario based on Higgs portal interactions. In Sec.~\ref{sec:results} we will provide sensitivity estimates of dark sector spectroscopy at the 14 and 100 TeV hadron collider and comment on the expected performance of a future lepton collider. As we will see, this will crucially depend on improved experimental systematics. In Sec.~\ref{sec:self-interaction} we connect the general discussion of Secs.~\ref{sec:model} and~\ref{sec:results} to the concrete case of self-interacting dark matter and demonstrate that aspects of dark sectors can in principle be revealed by studying high-energy collisions. In case a state will be discovered that can be interpreted as a dark sector-mediator, such measurements can complement the on-going effort to construct realistic composite dark matter scenarios using lattice simulations. We summarise and conclude in Sec.~\ref{sec:conc}.

%%%%%%%%%%%%%%%%%%%%%%%%%%
\section{Model}
\label{sec:model}
How can the dynamics of a strong sector influence the mediator phenomenology? The answer to this question is directly related to the UV properties of a particular mediator model and, to this end, we therefore focus on UV-complete models with scalar and vector mediators. As is well known from studies of simplified dark matter models~\cite{Frandsen:2012rk,An:2012va,Alves:2013tqa,Buchmueller:2013dya,Abdallah:2014hon,Buckley:2014fba,Harris:2014hga,Haisch:2015ioa,Jacques:2015zha,Khoze:2015sra,Abdallah:2015ter}, collider experiments are typically better suited to discover vector mediators with gauge-like interactions to quarks and the dark sector than scalar mediators with Yukawa-like couplings. In the vectorial case, however, the coupling to both visible and hidden sectors has to be a gauge coupling while the mass of the mediator is realised through spontaneous symmetry breaking (or a St\"uckelberg approach). The only possibility, therefore, is to understand the mediator interactions as part of a (Higgsed) product-group gauge theory, e.g. $\text{SM}\times U(1)_{\text{mediator}} \times SU(N)_\text{dark}$. Resumming the logarithmic enhancements of mediator production in e.g. a monojet signature can be estimated through a leading order (LO)-improved renormalisation group calculation that replaces the fixed LO value of $g'$ with running parameter as function of the probed energy scale. The behaviour of the mediator coupling $g'$ in the vectorial case, however, is protected through Ward identities which gives rise to a one-loop renormalisation group equation (RGE)
\begin{equation}
 \label{eq:gaugecoup}
 \mu \frac{\text{d}g'}{\text{d}\mu} \propto \frac{(g')^3}{16 \pi^2} \,,
\end{equation}
irrespective of the dynamics in either the visible or hidden sector. Moreover, the mediator production cross section would only reflect the total contributing number of degrees of freedom to the running of $g'$ but not their interaction properties (this is exactly the situation we encounter for the SM gauge couplings). On the one hand, such effects are difficult to observe in the LHC's (and a future 100 TeV collider's) energy range unless the value of $g'$ was large enough to make the validity of perturbation theory questionable and potentially introduce a low-scale Landau pole. On the other hand the monojet cross section dominantly probes the mediator sector only, which is not the question we would like to see addressed by the measurement.

The LO RGE characteristics of gauge couplings are not present for scalar mediators. This already becomes transparent from the SM RGEs, where the top Yukawa interaction behaves as
\begin{equation}
	\mu \frac{\text{d}y_t^{\text{SM}}}{\text{d}\mu} = \frac{y_t^{\text{SM}}}{ 16\pi^2} \left( \frac{9}{ 2} (y_t^{\text{SM}})^2 - \frac{17}{ 12} g_Y^2 -\frac{9}{ 4} g_L^2 - 8 g_3^2 \right)
\end{equation}
in addition to the RGE equation for the QCD coupling $g_3$ that now probes the strong dynamics as a consequence of Eq.~\eqref{eq:gaugecoup}. Therefore the combined solution of one-loop RGEs indeed dials sensitivity from the QCD sector into the behaviour of the Higgs-top interactions. This only happens at two-loop order for the gauge couplings and is, hence, suppressed in this case.

This shows that if we limit ourselves to scalar mediators, we can indeed expect to observe an echo of the strong sector dynamics in the mediator cross sections. Therefore, we focus in the following on a scenario consisting of a real SM-singlet scalar $\phi$ which obtains a vev $x$ (similar to the singlet-extended Standard Model~\cite{Binoth:1996au,Schabinger:2005ei,Patt:2006fw,Englert:2011yb,Baglio:2015wcg,Robens:2016xkb}), generating mass terms for three generations of SM-singlet Dirac fermion dark quarks $\psi$ through Yukawa interactions. These mass terms can be small while the heaviness of  the IR degrees of freedom can arise from confinement in the dark sector. The full scalar potential is given by\footnote{We impose a ${{Z}}_2$ symmetry to forbid any additional terms for simplicity.}:
\begin{multline}
{{ V(H,\phi) }}= - m_H^2 H^\dagger H - \frac{m_\phi^2}{2} \phi^2 + \lambda_1 (H^\dagger H)^2 \\ + \frac{\lambda_2}{4} \phi^4 + \frac{\lambda_3}{2} \phi^2 H^\dagger H.
\end{multline}
The $\lambda_3$ induced mixing between $\phi$ and $H$ generically results in interactions between the visible and dark sector mediated by the two scalar mass eigenstates $h$ and $h'$, and we denote the mixing angle $\theta$, defined through (in unitary gauge):
\begin{equation}
 H = \begin{pmatrix}0\\ (v + h_1)/{\sqrt{2}}\end{pmatrix}\,,\quad \phi =  (x + h_2)\,,
\end{equation}
which are related to the eigenstates in the Lagrangian by a two-dimensional isometry 
\begin{equation}
\begin{pmatrix} h \\ h' \end{pmatrix} = \begin{pmatrix}\cos \theta & -\sin\theta\\ \sin \theta & \cos \theta\end{pmatrix} \begin{pmatrix} h_1 \\ h_2 \end{pmatrix}.
\end{equation}
$\theta$ is expressed in terms of the Lagrangian parameters by
\begin{equation}
\label{eq:thetadef}
 \tan 2 \theta = \frac{ \lambda_3 v x}{\lambda_2 x^2 - \lambda_1 v^2}\,.
\end{equation}
In principle we have five free parameters in the scalar sector, which we choose as $m(h), m(h'),v,x,\theta$, but we identify $h$ as the Higgs-like particle discovered at the LHC which fixes $m(h) \simeq 125$ GeV and $v \simeq 246$ GeV.

%%%%%%%%%%%%%%%%%%%%%%%%%%%%%%%%
\begin{figure}[!t]
 \begin{center}
   \includegraphics[width=0.3\textwidth]{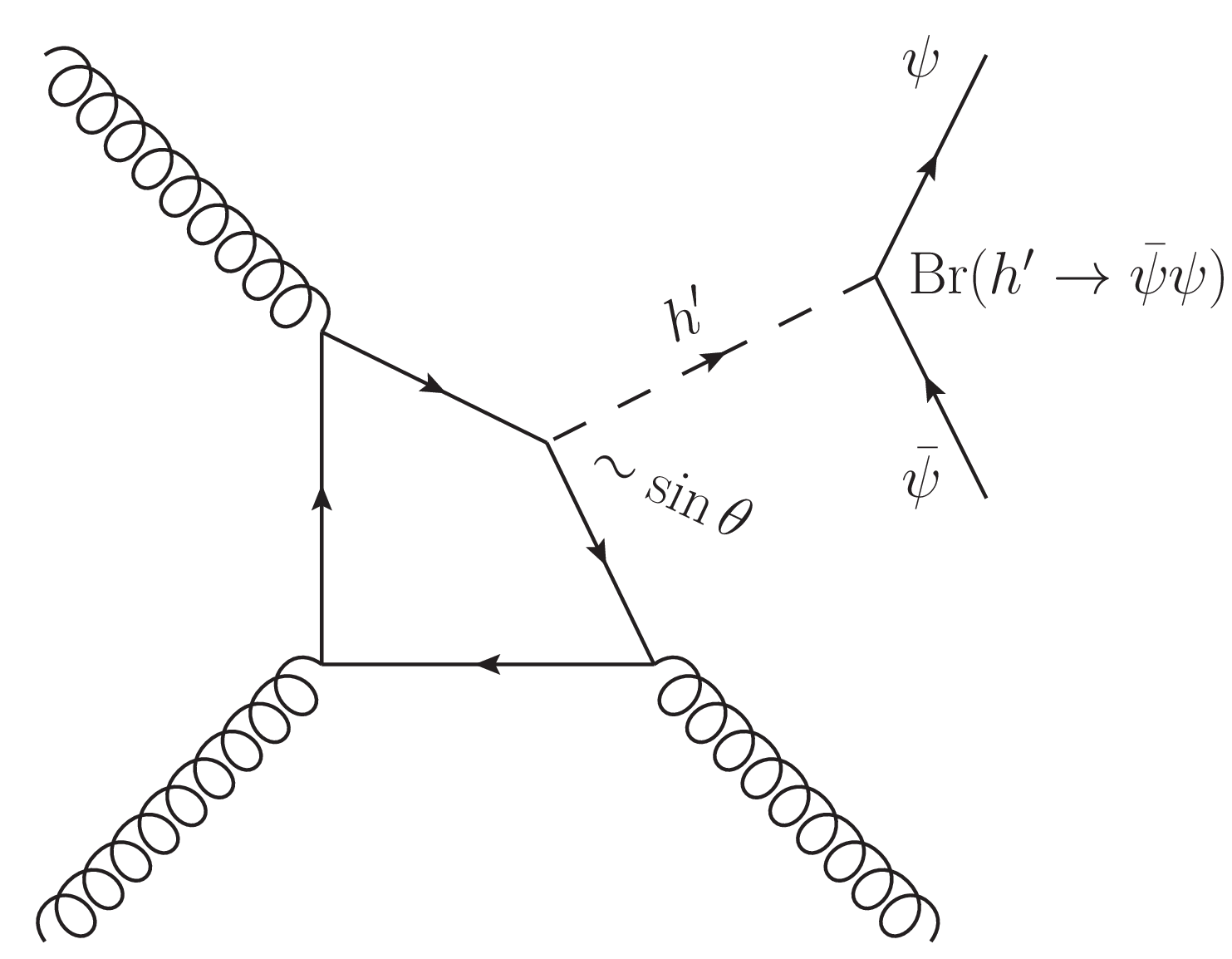}\\
 \end{center}
 \caption{\label{fig:diagram} Example diagram contributing to the process. The dependence on the dark gauge group enters through the running of $\theta$ as explained in the text. }
\end{figure}
%%%%%%%%%%%%%%%%%%%%%%%%%%%%%%%%% 

\subsection{The confining $SU(N)$ case}
Since we are interested in dark sectors with nontrivial gauge structure we introduce a new $SU(N)$ gauge group (under which the SM transforms as a singlet) and let the dark quarks transform in the $\textbf{M}$ representation. We consider the $U(1)$ case below. 

The dark sector Lagrangian then reads
\begin{equation}
 \mathcal{L}_{\text{dark}} = -Y_\psi^{i,j} \phi \bar{\psi}^i \psi^j + \text{h.c.} + i \bar{\psi} \gamma^\mu D_\mu \psi - \frac{1}{4g_d^2} G_{\mu \nu}^a G^{a, \mu \nu} \,.
\end{equation}
The mixing of the scalars will modify the interaction strength of the mass eigenstates with the two matter sectors: $h$ couplings to the Standard Model are scaled by $\cos \theta$ compared to the Standard Model expectation and dark sector couplings are scaled by $\sin \theta$ compared to the $\phi$, and vice versa for $h'$. This means $\bar{\psi} \psi$ production is allowed through both $h$ and $h'$ when kinematically possible (see~Ref.~\cite{Dupuis:2016fda} for a more detailed look at the phenomenology of a similar model). We assume $Y_\psi$ is diagonal which means we have four new parameters but motivated by the structure of the Yukawa terms in the SM we assume the third generation of dark quarks is considerably heavier than the two first ones and set the other Yukawa terms to 0, which leaves us with $Y_\psi^{3,3}$ and $g_d$. Showering and hadronisation can then occur as in QCD~\cite{Strassler:2006im,Han:2007ae} and decay to low-lying states can be achieved through additional weak interactions which will not impact the qualitative scaling behavior induced by the strong interactions in the dark sector (like in the SM sector). In total our free parameters for our study are
\begin{equation}
 Y_\psi^{3,3}, g_d, \theta, m(h'), x \,.
\end{equation}
To illustrate the effects of RGE running from different $N$ and $\textbf{M}$ we fix most of these to generic values inspired by their SM equivalents (defined at the $h$ pole): $x = 100$ GeV, $Y_\psi^{3,3} = 0.7$, $\theta = 0.5$, and $m(h') = 150$ GeV. This parameter point is in agreement with current constraints~\cite{Robens:2015gla}. For our chosen benchmark of 70 GeV fermion mass production through $h$ is kinematically suppressed and we will ignore it from now on. Also, since $\text{Br}(h' \to \psi \bar{\psi}) \approx 1$ current constraints from additional Higgs searches in visible channels are easily evaded.

We fix $g_d$ in two different ways: first, by setting $g_d = g_S$ at the $Z$ pole in order to map out the general features of the solutions in section~\ref{sec:results}, and second, by requiring the dark IR Landau pole to be $\sim$0.5 GeV in order to make $\Lambda_d$ fall in a relevant part of parameter space for self-interacting dark matter in section~\ref{sec:self-interaction}. This second requirement could be refined by using auxiliary measurements (e.g. on the lattice) but should capture the main features we are interested in; relevant to our analysis is the comparison of the different dark sectors.

%%%%%%%%%%%%%%%%%%%%%%%%%%%%%%%%%
\begin{figure*}[!t]
 \begin{center}
  \subfigure[\label{fig:sun} ]{\includegraphics[width=0.43\textwidth]{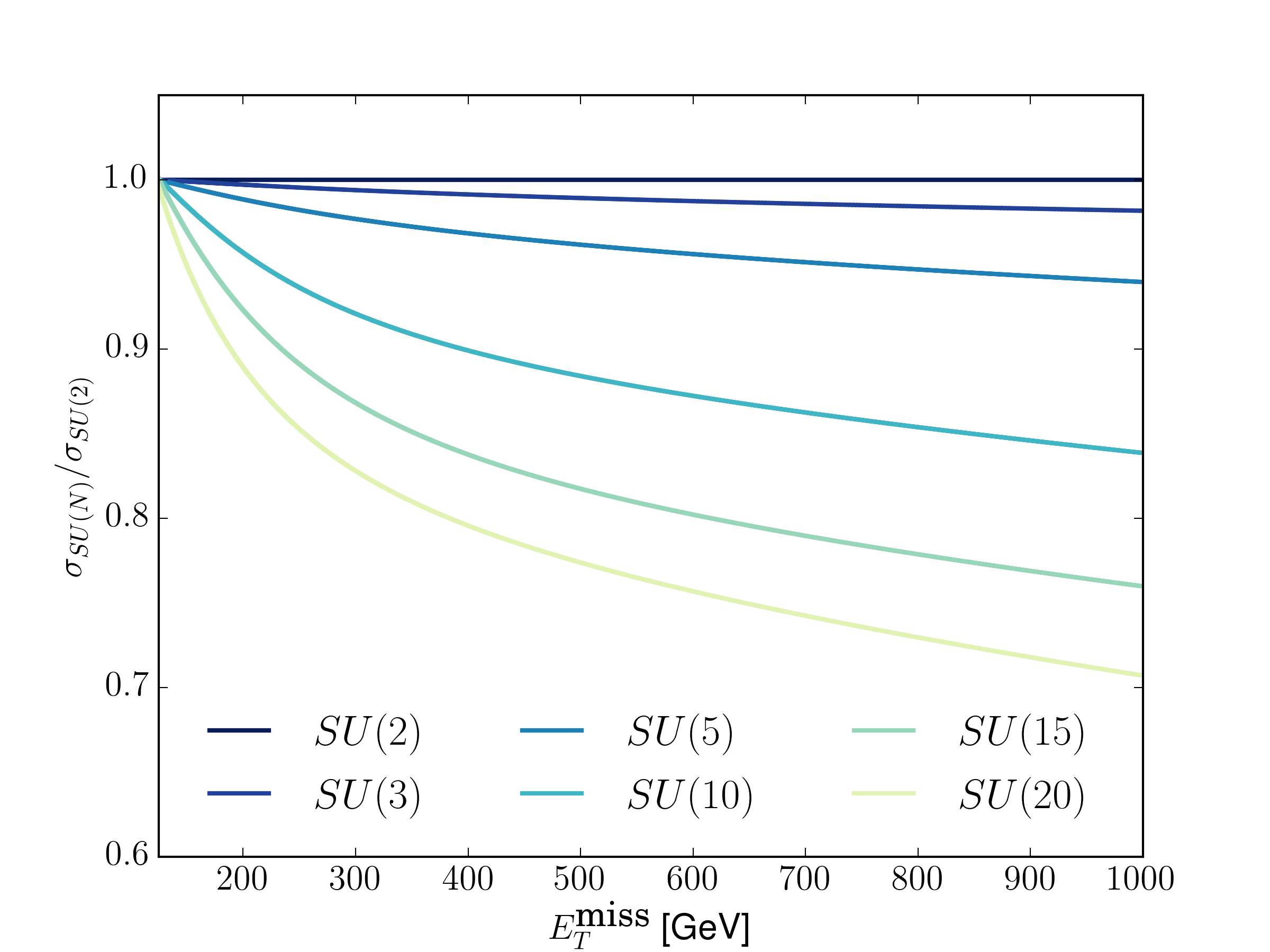}}\hspace{1cm}
  \subfigure[\label{fig:repsu5} ]{\includegraphics[width=0.43\textwidth]{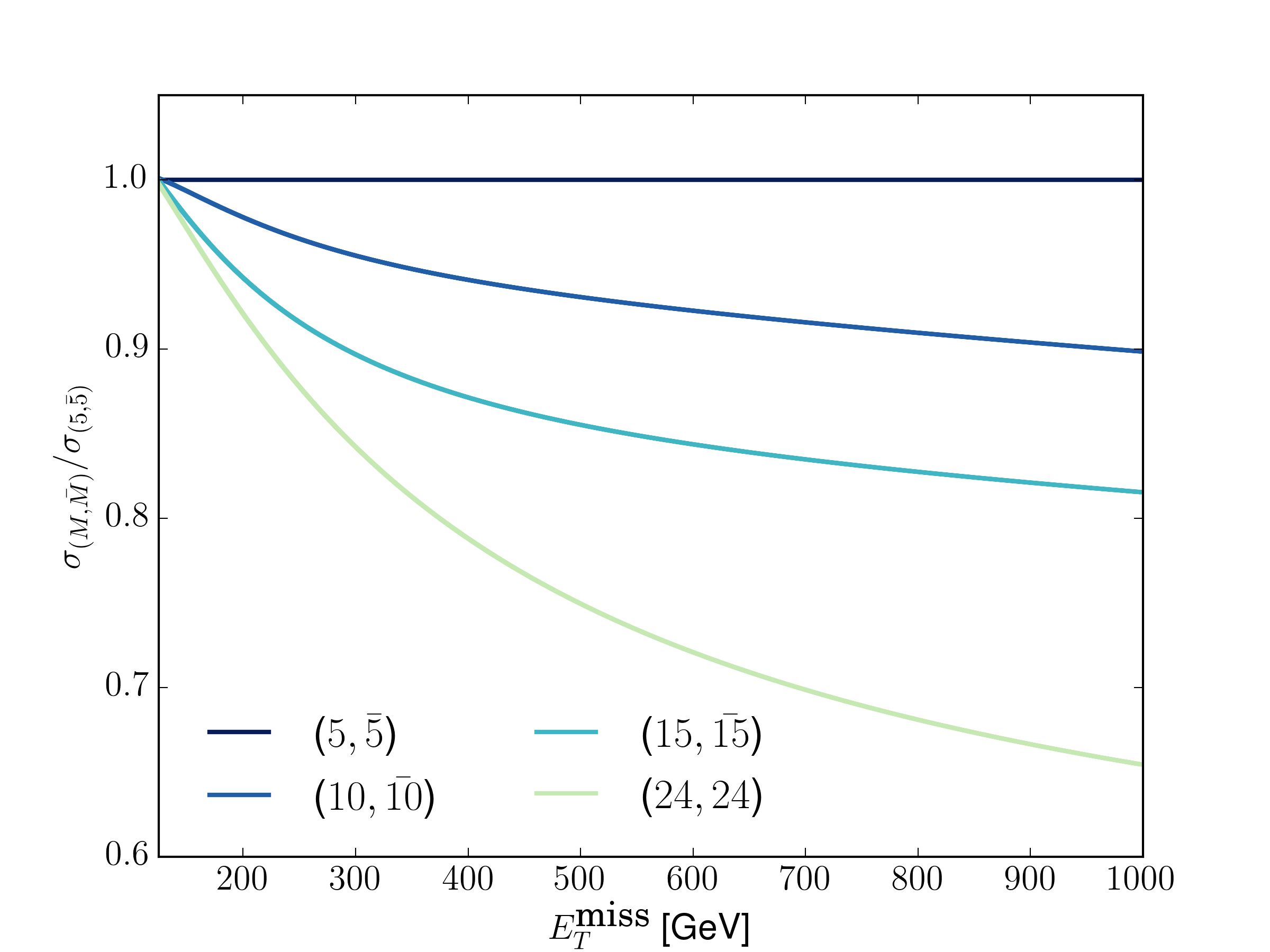}}\\ 
  \vspace{-0.4cm}
  \subfigure[\label{fig:u1} ]{\includegraphics[width=0.43\textwidth]{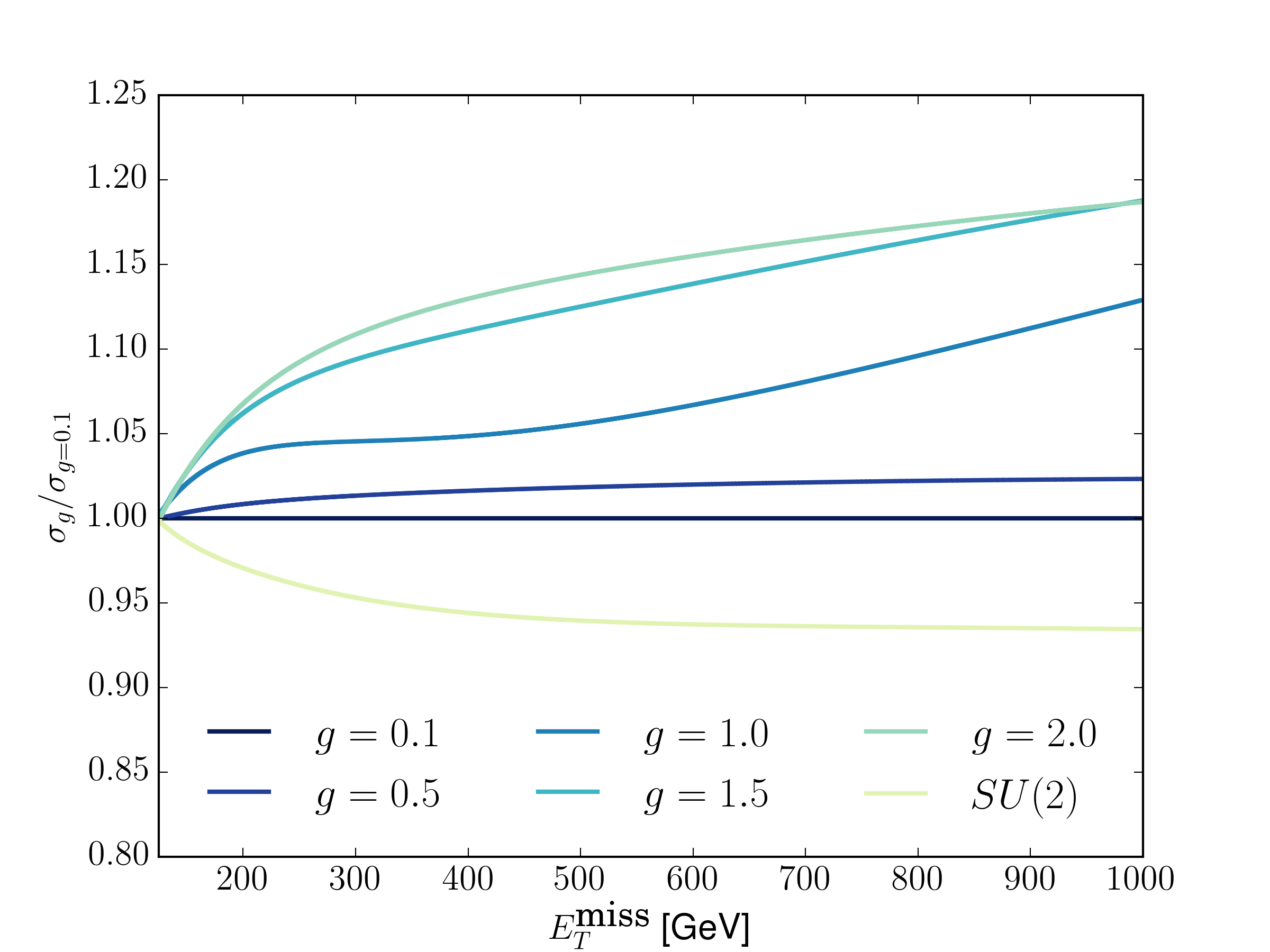}}\hspace{1cm}
  \parbox{0.43\textwidth}{\vspace{-4cm}\parbox{0.37\textwidth}{
  \caption{\label{fig:spectra} Ratios of monojet channel cross sections in the missing $E_T^{\text{miss}}$ spectra for~\ref{fig:sun} varying gauge groups
   with the dark quarks in the fundamental representation, \ref{fig:repsu5} for $SU(5)$ with the dark
   quarks in varying representations, and~\ref{fig:u1} for the $U(1)$ model with varying values for the dark coupling $g$ fixed at $M_Z$. We fix $g_d = g_S$ at $M_Z$ as the dark gauge sector boundary condition.}
   }}
 \end{center}
\end{figure*}
%%%%%%%%%%%%%%%%%%%%%%%%%%%%%%%%% 

Much like the top Yukawa in the Standard Model, the $\beta$ function of $Y_\psi^{3,3}$ will be sensitive to the dark gauge group already at one-loop level, which is the source of the dependence on the precise form of the group of the mixing angle $\theta$ in Eq.~(\ref{eq:thetadef}), as the one-loop $\beta$ functions for $\lambda_2$, $\lambda_3$ and $x$ all have a dependence on $Y_\psi^{3,3}$. Additionally these also depend on $\textbf{M}$~\cite{MACHACEK198383,MACHACEK1984221,Luo:2002ti} 
\begin{widetext}
\begin{subequations}
\begin{alignat}{5}
\label{eq:gd}
  \mu \frac{\text{d}g_d}{\text{d}\mu} &= - \left( \frac{11}{3} C(\text{\textbf{A}}) - 4 \hspace{0.1cm} T(\textbf{M}) \right) \frac{g_d^3}{16 \pi^2}\,, \\
\label{eq:ypsi}
\mu \frac{\text{d}Y_\psi^{3,3}}{\text{d}\mu} &=\left( -  6 \hspace{0.1cm} C(\text{\textbf{M}}) \hspace{0.1cm} g_d^2 + (2 \text{ Dim(\textbf{M})} + 3) (Y_\psi^{3,3})^2  \right) \frac{Y_\psi^{3,3}}{16 \pi^2}\,,\\
\label{eq:lam2}
  \mu \frac{\text{d}\lambda_2}{\text{d}\mu} &= \left( 18 \lambda_2^2 + 2 \lambda_3^2 + 8 \text{ Dim(\textbf{M}) } \lambda_2 (Y_\psi^{3,3})^2 - 8 \text{ Dim(\textbf{M}) } (Y_\psi^{3,3})^4 \right) \frac{1}{ 16 \pi^2}\,,
\end{alignat}
\begin{alignat}{5}
\label{eq:lam3}
 \mu \frac{\text{d}\lambda_3}{\text{d}\mu} &= \left(-\frac{3}{2} g_1^2 - \frac{9}{2} g_2^2 + 12 \lambda_1 + 6 \lambda_2  + 4 \lambda_3   + 4 \text{ Dim(\textbf{M}) } (Y_\psi^{3,3})^2 + 6 y_t^2 \right)  \frac{\lambda_3}{16 \pi^2} \,,\\
\label{eq:x}
 \mu \frac{\text{d}x}{\text{d}\mu} &= - 2 \text{ Dim(\textbf{M}) } (Y_\psi^{3,3})^2 \frac{x}{16 \pi^2}\,.
\end{alignat}
\end{subequations}
\end{widetext}
Here $C(\text{\textbf{A}})$ is the quadratic Casimir of the adjoint representation ($= N$), $T(\textbf{M})$ is the index of $\textbf{M}$, and $C(\text{\textbf{M}})$ the quadratic Casimir of $\textbf{M}$.

Taking the Standard Model as a guiding example, it is also reasonable to expect an $SU(N)$ to capture the most important RGE effects even when the gauge group is enlarged, and hence this study should have some applicability beyond the simple scenario we consider here.

\subsection{Dark $U(1)$s}
We also consider a model with a dark $U(1)$ symmetry which the now complex $\phi$ is charged under, hence generating a mass term for the new gauge boson using the extra scalar degree of freedom. To avoid anomalies we have to introduce an additional dark fermion field and we choose the charges as $q_{d} \sim 0$, $u_{d} \sim 1/2$, $d_{d} \sim -1/2$, $\phi \sim 1/2$. The dark sector Lagrangian is then:
\begin{multline}
\mathcal{L}_{\text{dark}} = -Y_u^{i,j} \phi^\dagger \bar{u}_{d}^i q_d^j - Y_d^{i,j} \phi \bar{d}_{d}^i q_d^j + \text{h.c.} \\ + i \bar{q}_{d} \gamma^\mu D_\mu q_d 
+ i \bar{u}_{d} \gamma^\mu D_\mu u_d + i \bar{d}_{d} \gamma^\mu D_\mu d_d - \frac{1}{4g^2} F_{\mu \nu} F^{\mu \nu}\,.
\end{multline}
This gives us a theory which is similar to the one introduced above but which is not confining and has a Yukawa-like interaction potential between the dark fermion fields. Note that we will refer to the gauge coupling in the $U(1)$ model as $g$ in contrast to $g_d$ in the non-Abelian case. Much like in the non-Abelian case we assume only the heaviest fermion with a mass of 70 GeV is relevant for our RGE calculation and set all other Yukawa terms to 0, and keep all other parameters the same. We also assume there is no kinetic mixing between the dark $U(1)$ and $U(1)_Y$. The renormalisation group equations for this model are given below. Note that we have changed the normalisation of the $\phi$ field to that of a complex scalar field, and include a factor of $1/\sqrt{2}$ when expanding around $x$ (leading to factor of 2 difference for terms involving squared Yukawas).

The RGEs for this model read~\cite{MACHACEK198383,MACHACEK1984221,Luo:2002ti} 
\begin{widetext}
\begin{subequations}
\begin{alignat}{5}
\label{eq:g_u1}
  \mu \frac{\text{d}g}{\text{d}\mu} &= \frac{13}{12} \frac{g^3}{16 \pi^2}\,, \\
\label{eq:yu_u1}
\mu \frac{\text{d}Y_u^{3,3}}{\text{d}\mu} &=\left( -\frac{3}{4} g^2 + 2 (Y_u^{3,3})^2   \right) \frac{Y_u^{3,3}}{16 \pi^2}\,,\\
\label{eq:lam2_u1}
  \mu \frac{\text{d}\lambda_2}{\text{d}\mu} &= \left( 20 \lambda_2^2 + 2 \lambda_3^2 + \frac{3}{8} g^4 - 3 g^2 \lambda_2 +4 \lambda_2 (Y_u^{3,3})^2 - 2(Y_u^{3,3})^4 \right) \frac{1}{ 16 \pi^2}\,,\\
\label{eq:lam3_u1}
 \mu \frac{\text{d}\lambda_3}{\text{d}\mu} &= \left(-\frac{3}{2} g_1^2  - \frac{9}{2} g_2^2 -\frac{3}{2} g^2 + 12 \lambda_1 + 8 \lambda_2  + 4 \lambda_3   + 2 (Y_u^{3,3})^2 + 6 y_t^2 \right)  \frac{\lambda_3}{16 \pi^2} \,,\\
\label{eq:x_u1}
 \mu \frac{\text{d}x}{\text{d}\mu} &= -  (Y_u^{3,3})^2 \frac{x}{16 \pi^2}\,.
\end{alignat}
\end{subequations}
\end{widetext}

\section{Results}
\label{sec:results}
We use \textsc{Sarah}~\cite{Staub:2013tta} in order to obtain the relevant $\beta$ functions at one loop for the described scenarios (checked against the general forms given in~\cite{Luo:2002ti}), and solve these for the given boundary conditions.\footnote{\textsc{Sarah} also calculates the two-loop $\beta$ functions on demand but we do not use these to keep the dependence on $N$ and $\textbf{M}$ completely transparent as detailed in (\ref{eq:gd})-(\ref{eq:x}). We have checked that including two-loop effects does not change the results presented here.} These are then used to calculate the running of the mixing angle, which is then passed on to a \textsc{FORTRAN} implementation of a full leading-order $p p \to h' j$ parton-level event generator based on \textsc{Vbfnlo}~\cite{Arnold:2011wj} and \textsc{FormCalc}~\cite{Hahn:1998yk,Hahn:2000kx}, arriving at a one-loop RGE-improved parton-level calculation which is typically used in QCD LO calculations. Finally we take the branching ratio $\text{Br}(h' \to \psi \bar{\psi})$ into account as a flat rescaling at the Higgs masses, which corresponds to the advocated prescription of the Higgs Cross Section Working Group~\cite{Heinemeyer:2013tqa}. Throughout, the scale in the calculation is set to $p_T(h')$, which is a motivated relevant scale for the logarithmically enhanced modifications of the cross section at large momentum transfers.

%%%%%%%%%%%%%%%%%%%%%%%%%%
\squeezetable
\begin{table}[!b]
\begin{tabular}{ l || c | c | c | r }
  Cut & $U(1)$ & $SU(2)$ & $SU(25)$ & Bgd. \\
  \hline
  $E_T^\text{miss}>$ 200 GeV & 1.84 pb & 1.70 pb & 1.45 pb & 432 pb \\
  $E_T^\text{miss}>$ 500 GeV & 0.0411 pb &  0.0359 pb & 0.0271 pb & 18.0 pb \\
  \hline
  signal Ratio		& $44.8 \pm 1.47$ & $47.3 \pm 1.78$ & $53.5 \pm 2.66$ & 
\end{tabular}
\caption{\label{tab:100tev} Cross sections of the signal at 100 TeV and expected measurements of the scaling with $E_T^\text{miss}$ using 10 ab$^{-1}$ of data. The $U(1)$ result uses $g(M_Z) = 0.1$. The statistics-only uncertainty on the ratio is calculated by estimating the statistical uncertainty on the signal strength in both cases and propagating these through to the ratio. For a CLs test based on the missing energy distribution see below.}
\end{table}
%%%%%%%%%%%%%%%%%%%%%%%%%%

\subsection*{14 and 100 TeV hadron colliders}

We estimate the monojet background by generating $pp \to (Z \to \nu \nu) j$ parton-level events and scaling this by a factor of 1.5 to get an estimate of the total background following~Ref.~\cite{Aad:2015zva}.\footnote{While mismeasured lepton $(W\to \nu l)j$ events are important and slightly change the scaling with energy of the background, this rescaling should be conservative for our purposes as backgrounds at larger $E_T^\text{miss}$ will be over-estimated.}

In order to get a handle on the strong sector dynamics, we need to study the energy dependence of exclusive cross sections. Concretely this means we need to determine how an excess in the monojet channel scales as a function of $E_T^{\text{miss}}$ when such a signal can be extracted from the background. This will allow us to make a statement about the likely gauge structure of the dark sector if different dark gauge groups indeed predict a statistically relevant deviation in a comparison. The relative scaling of the cross section as a function of missing $E_T^{\text{miss}}$ for different gauge groups and different representations of $SU(5)$ is given in Fig.~\ref{fig:spectra}. Due to our choice of scale the behaviour will be exactly the same at all center-of-mass energies.

The constraints from single Higgs phenomenology enforce a small mixing angle for SM-like Higgs measurements, which act as a boundary condition to the RGE flow. We therefore find for our parameter point that the absolute cross sections at 14 TeV are too small for a measurement to be made even with the full HL-LHC data set.\footnote{By changing the parameters, however, we could indeed maximise the potential of the HL-LHC at the price of creating further tension with Higgs signal strength measurements. We do not discuss this case in detail as it is likely to be challenged by run 2 analyses.}

%%%%%%%%%%%%%%%%%%%%%%%%%%%%%%%%%
\begin{figure}[!t]
 \begin{center}
 \includegraphics[width=0.43\textwidth]{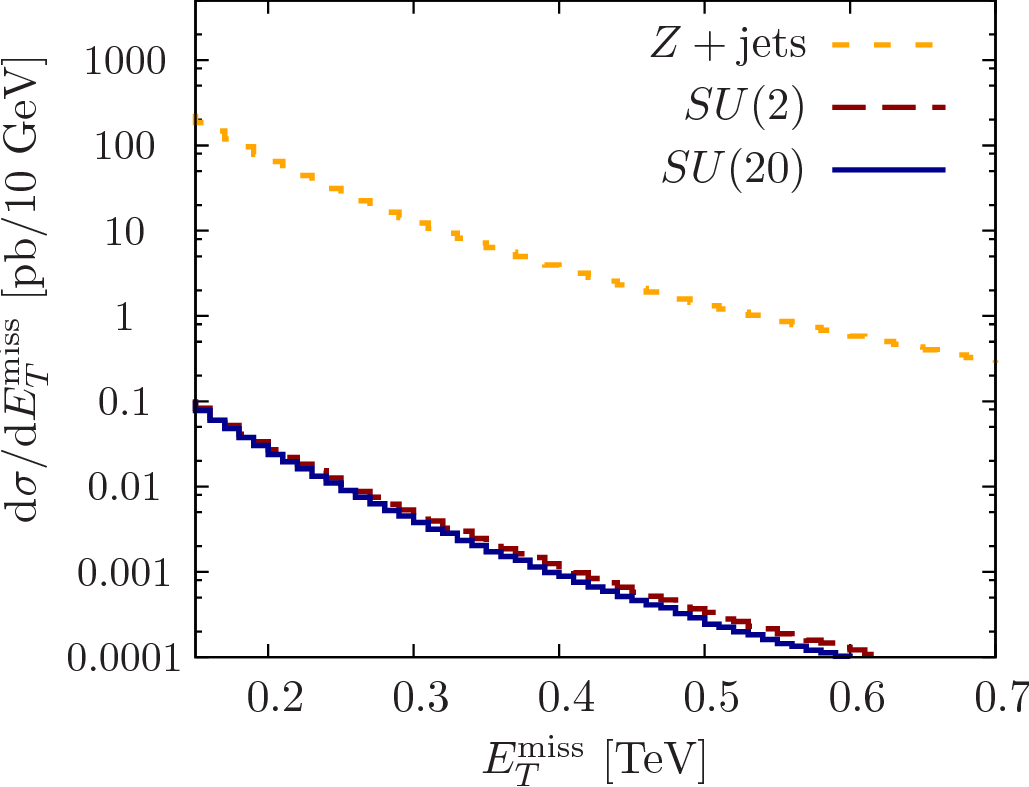}
 \end{center}
 \caption{\label{fig:spectraetmiss} 100 TeV signal and background distributions that feed into the confidence level calculation detailed in the text.}
\end{figure}
%%%%%%%%%%%%%%%%%%%%%%%%%%%%%%%%% 

The cross sections at a 100 TeV proton-proton collider given in Table~\ref{tab:100tev} are large enough to offer an opportunity to make a measurement of the running of $\theta$ using a data set of 10 ab$^{-1}$. Given the expected small mixing angle, the largest experimental challenge will undoubtedly be the reduction of the systematic uncertainties of the measurements by over an order of magnitude compared to the recent 8 and 13 TeV monojet analyses by ATLAS and CMS~\cite{Aad:2015zva,Khachatryan:2014rra,Aaboud:2016tnv,CMS:2016tns}. The impeding factor of a 14 TeV analysis, i.e. the smallness of the expected signal cross section as well as a limited data set will be overcome at a 100 TeV machine, where the signal cross sections are large enough to gather very large statistics with the aim to use data-driven, as well as multivariate techniques, which essentially remove the background uncertainties to a very large extent. Using an extrapolation from the low-missing-energy regime is not straightforwardly possible since the low-missing-energy phase space region receives a contribution from signal events, and is not entirely background dominated. However, $Z$ boson data can be extrapolated from visible $Z\to e^+e^-$ and $\gamma+\text{jet}$ subsidiary measurements at essentially zero statistical uncertainty (note that all involved couplings are gauge couplings following \eqref{eq:gaugecoup}), which essentially allows us to directly infer the dominant $Z(\to \nu\bar \nu)+\text{jet}$ distribution completely using data-driven techniques. Similar techniques were used already for 8 TeV analyses, e.g.~\cite{CMS:2015dbr} (see also~\cite{Englert:2011cg,Bern:2011pa,Gerwick:2012hq,Bern:2014fea,Mangano:2015ejw,Bothmann:2016loj} for related theoretical work). Since the detector layout of a 100 TeV machine is likely to change towards an improved electromagnetic calorimeter coverage~\cite{Golling:2016gvc,Arkani-Hamed:2015vfh}, this mapping from $(Z\to e^+e^-)+\text{jet}$ and $\gamma+\text{jet}$ could also be performed without relying on an extrapolation into the jet-acceptance region beyond the lepton and photon acceptance regions $|\eta|<2.5$ that is imposed by the current LHC setup.

%%%%%%%%%%%%%%%%%%%%%%%%%%%%%%%%%
\begin{figure}[!t]
 \begin{center}
 \includegraphics[width=0.43\textwidth]{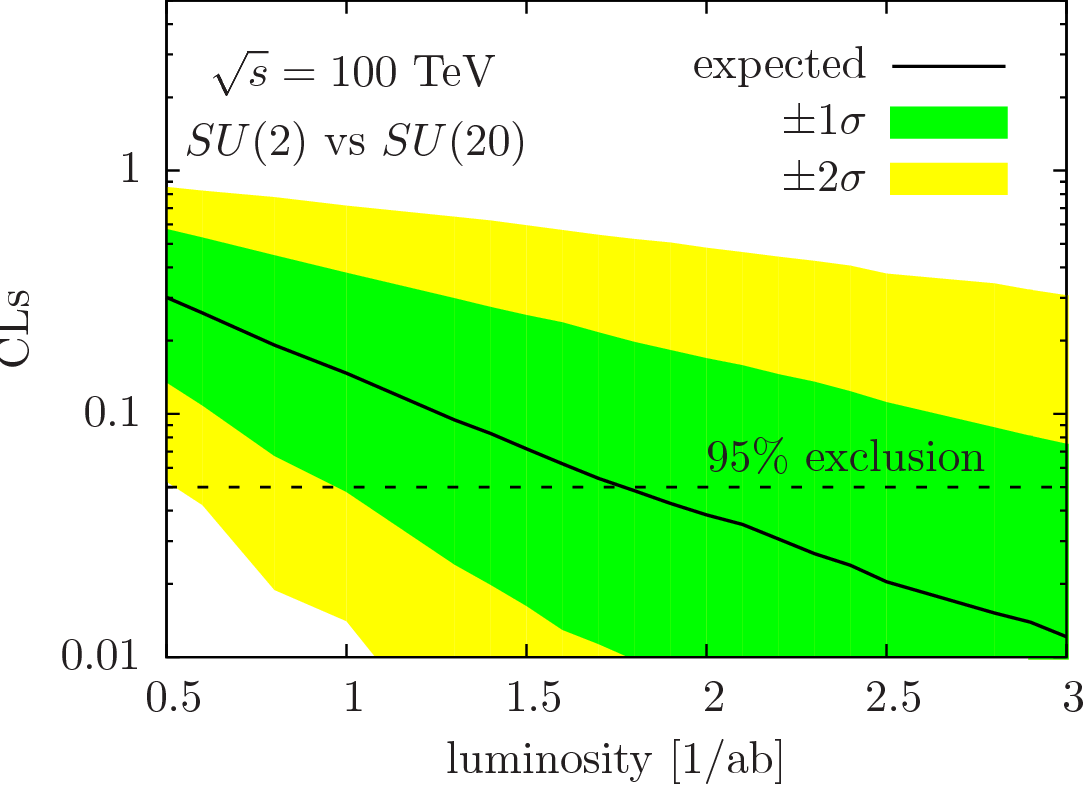}
 \end{center}
 \caption{\label{fig:cls} CLs hypothesis test detailed in the text, only assuming statistical uncertainties.}
\end{figure}
%%%%%%%%%%%%%%%%%%%%%%%%%%%%%%%%% 

 In the likely case that we can gain excellent control over the background distribution in a data-driven approach (i.e. assuming only statistical uncertainties), we can expect a $5\sigma$ discovery threshold of $\gtrsim 100~\text{fb}^{-1}$ using a binned log-likelihood approach (as detailed in Refs.~\cite{Junk:1999kv,James:2000et}) based on the missing energy distribution for the $SU(2)$ running, although the signal vs. background ratio is small. Discriminating the $SU(2)$ from the $SU(20)$ hypothesis, for instance, should then be possible at 95\% CLs~\cite{Read:2002hq} for ${\cal{L}}\gtrsim 1.6~\text{ab}^{-1}$ (assuming statistical uncertainties only, Fig.~\ref{fig:cls}). Similar conclusions hold for discriminating large non-Abelian groups against the $U(1)$ scenario (at slightly smaller integrated luminosities).
 
\subsection*{Probing dark sectors through $h$ couplings}

Since our SM-like scalar $h$ has its couplings scaled by $\cos(\theta)$ we could in theory also use these to investigate the structure of the dark sector. Measuring $\theta$ at $m_h$ is straightforward (current measurements already put some tension on our parameter point), after which the scaling could be investigated through a similar analysis as above but using cleaner and better understood visible decay channels, with larger cross sections. However, the issue with such a measurement is that $\theta$ generically runs to smaller values and hence towards the maximum of $\cos\theta$ where derivatives vanish; at small $\theta_{1,2}$, $\cos(\theta_1)/\cos(\theta_2) \sim 1$. At our parameter point we find cross section differences of up to 30\% between $SU(2)$ and $SU(20)$ at a scale of 1 TeV when looking at production scaled by $\sin(\theta)^2$ (running away from the minimum), but these differences shrink to about 1\% when scaling by $\cos(\theta)^2$. At smaller values of $\theta(m_h)$ this problem is further worsened and already with $\theta(m_h) = 0.1$ one needs to investigate differences of $\mathcal{O}(0.01\%)$, which is challenging the sensitivity range of a future lepton collider~\cite{Peskin:2012we,Blondel:2012ey}.

\subsection*{A note on future lepton colliders}
At a future lepton collider the two dominant production mechanisms for $h'$ would be $h'$-strahlung and $WW$ fusion. $h'$-strahlung is a threshold effect and would as such be inherently insensitive to the running of $\theta$. $WW$ fusion dominates at higher $\sqrt{s}$ and does in theory feel the running of $\theta$ but since the final state would be $h' \nu \bar{\nu}$, a measurement would have to rely on a radiated photon leading to cross sections of the order of $O(1-10 \text{ fb})$ for $\sqrt{s} = 500 - 1000$ GeV for unpolarised $e^+ e^-$ beams, making a measurement dependent on extremely large integrated luminosities. However, thanks to the controlled kinematics at a lepton collider, the dominant background $(Z \to \nu \bar{\nu}) \gamma$ peaks strongly at $E_\gamma = E_Z = E_\text{beam}$ for $E_\text{beam} \gg m_Z$, whereas the signal peaks below $E_\gamma < m_{h'}$, which allows for an almost background-free analysis before detector effects are taken into account. The choice of scale here is not straightforward and we can expect non-RGE electroweak effects to play a significant role. Although this channel provides a clean avenue to test the hypothesis, RGE analyses alone cannot obtain a reliable estimate of the sensitivity.

\section{Potential Relation with self-interacting dark matter}
\label{sec:self-interaction}
The DM self-interaction cross section is measured at a very large and mass-dependent value $\sigma/m\simeq 1.3~\text{b/GeV}$~\cite{Randall:2007ph}. Such large cross section can of course be achieved by going to very small mass scales in the perturbative regime (see e.g.~Ref.~\cite{Ko:2014nha}) and our U(1) discussion of the previous section is therefore directly relevant for these scenarios. 

Cross sections of this size are not unusual in strongly interacting confined theories such as QCD, and we focus on this possibility in the following in detail. While the non-Abelian theories we have discussed so far are asymptotically free, explaining the relatively large characteristic decrease at large momentum transfers, they will confine at low scales to giving rise to a series of hadronic states in the dark sector. The details are highly dependent on the respective fermion and gauge symmetry content and the details as well as the existence of realistic confining theories can only be clarified by lattice simulations. However, we can obtain a qualitative estimate of whether such theories can reproduce self-interacting dark matter scenarios by means of chiral perturbation theory ($\chi$PT). To this end we assume that the self-interaction cross section is dominated by nonrelativistic pion scattering, well below the energy scales of other dark hadronic resonances. This will provide an estimate of the validity range of such scenarios and give us an idea if our previous discussion is relevant for self-interacting dark matter scenarios without making a particular reference to modified velocity distributions of the dark matter halo, which are likely to be found in theories with complex interactions~\cite{McCullough:2013jma,Fan:2013tia,Fan:2013tia}. Modifications from both corrections due to additional hadronic contributions to the cross section as well as a modified dark matter profile will change our numerical outcome, but can be compensated at least numerically by changing the fundamental parameters of $\chi$PT, which needs to be confirmed by lattice investigations. 

\begin{figure}[!t]
 \begin{center}
  \subfigure[\label{fig:sun_landau} ]{\includegraphics[width=0.43\textwidth]{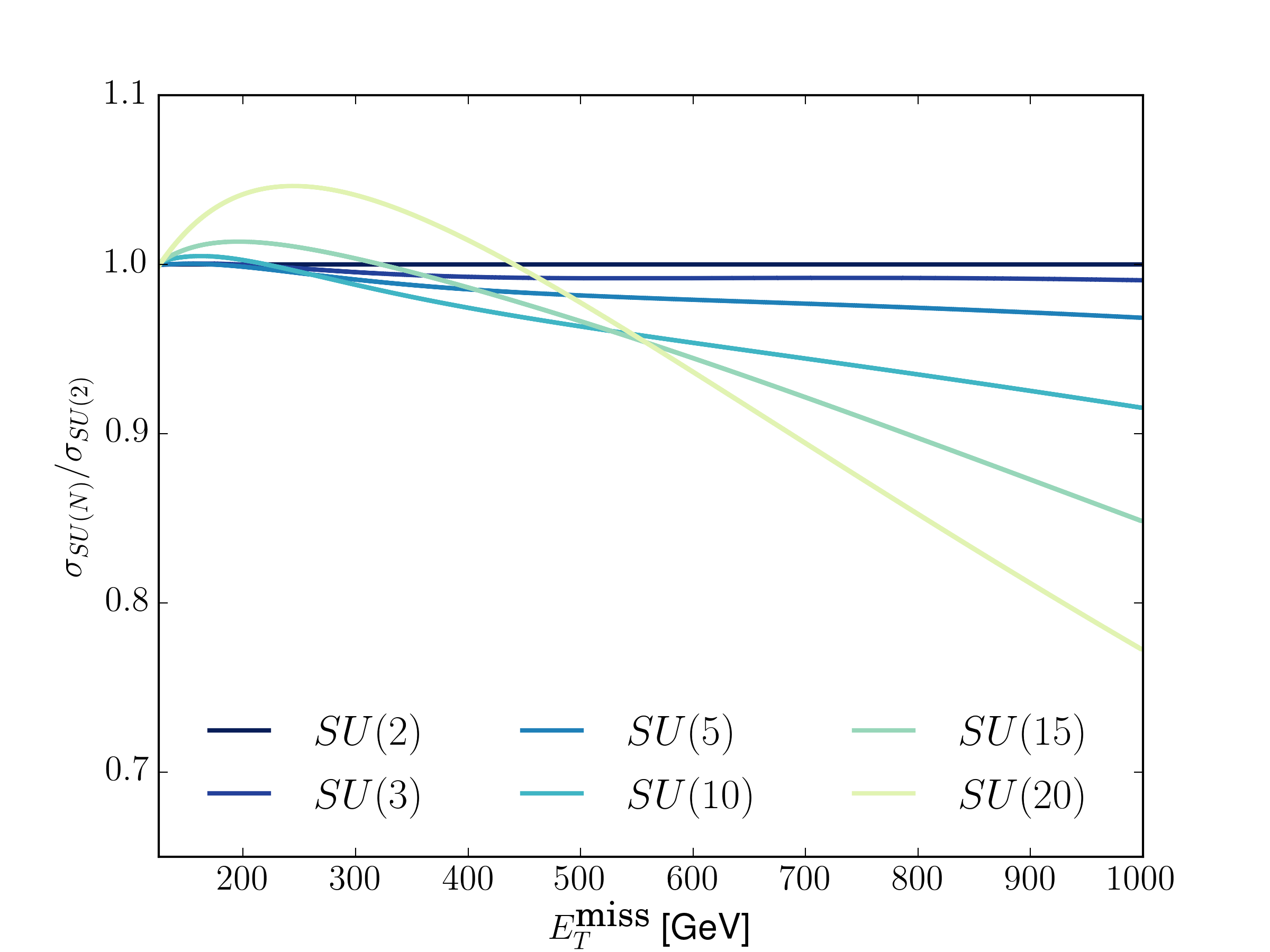}}\\
  \vspace{-0.4cm}
  \subfigure[\label{fig:repsu5_landau} ]{\includegraphics[width=0.43\textwidth]{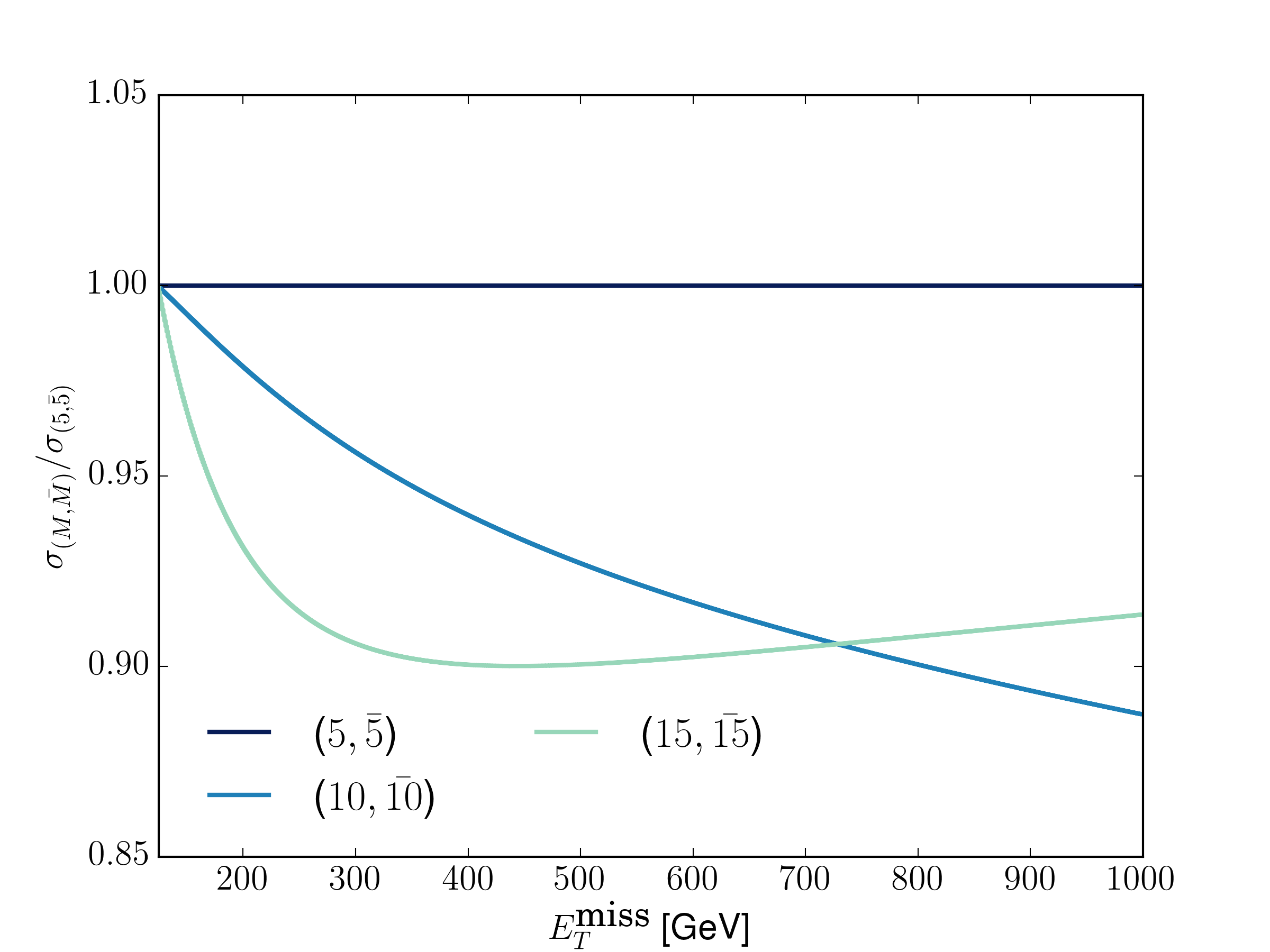}}\\
  \vspace{-0.4cm}
 \end{center}
 \caption{\label{fig:spectra_landau} Ratios of monojet channel cross sections in the missing $E_T^{\text{miss}}$ spectra for~\ref{fig:sun_landau} varying gauge groups
   with the dark quarks in the fundamental representation, \ref{fig:repsu5_landau} for $SU(5)$ with the dark
   quarks in varying representations. The value of $g_d$ was fixed by requiring $\Lambda_d \simeq 0.5$ GeV. }
   \vspace{-0.4cm}
\end{figure}

The pion dynamics is completely determined by a $[SU(N)\times SU(N)]/SU(N)$ nonlinear sigma model describing the coset field $\Phi(x)$ with dark pion decay constant $f_{\tilde{\pi}}$
\begin{equation}
	U(x)=\exp\left( \frac{i \Phi (x) }{ f_{\tilde{\pi}}} \right)\,.
\end{equation}
Analogous to QCD we assume the pion to be the lightest hadronic state in the spectrum; if no additional gauged $U(1)$ symmetry is present in the dark sector this state will remain stable.\footnote{Quasi-singularities exist which could potentially lead to a large \textit{CP} violation effect through interactions mediated by 't Hooft vertices. These involve coherent dark quark fields which are difficult to maintain at high temperatures in the early Universe and therefore SM baryogenesis is hard to explain by communicating dark baryogenesis to the visible sector.}
The interactions that we consider follow from expanding the non-linear sigma model
\begin{equation}
{\cal{L}}_{\text{dark},\chi}= \frac{f_{\tilde{\pi}}^2}{ 4} \text{Tr} \left(\partial_\mu U \partial^\mu U^\dagger \right)
\end{equation}
and by identifying the dark pion with the uncharged pion analogous to QCD we arrive at
\begin{equation}
{\cal{L}}_{\text{dark},\chi} = \frac{1}{ 2} (\partial\pi)^2 + \frac{1}{ {f_{\tilde{\pi}}^2}} \pi^2 (\partial \pi )^2 + \dots\,,
\end{equation}
where the ellipsis refers to higher-order terms in the $\chi$PT expansion as well as interactions of other states. With this Lagrangian we can compute the self-interaction cross section straightforwardly (we have cross-checked our results against implementations with {\sc{FeynRules}}~\cite{Alloul:2013bka} and {\sc{FormCalc}}~\cite{Hahn:1998yk,Hahn:2000kx}) and obtain in the nonrelativistic limit
\begin{equation}
\label{eq:chixs}
	\frac{\sigma}{ m}=\frac{m }{ 4\pi{f_{\tilde{\pi}}^4}}\,,
\end{equation}
which we can use to gauge whether self-interaction cross sections can be obtained from theories that show similarities with QCD (we assume the mass to be generated through a small explicit chiral symmetry violation analogous to QCD). With naive dimensional analysis~\cite{Manohar:1983md} (NDA), we can furthermore limit the parameter range of the dark pion decay constant given its mass. The mass needs to be smaller than the NDA cutoff $m<\Lambda_{\text{dark}} \simeq 4\pi f_{\tilde{\pi}}$ and pion scattering needs to be in agreement with the observed self-interaction cross section of $\sigma/m\simeq 1.3~\text{b/GeV}$. This locates the cutoff of the theory between $0.2~\text{GeV}\lesssim 4\pi f_{\tilde{\pi}} \lesssim 0.8~\text{GeV}$ for pion masses $m<0.8~\text{GeV}$. Matching the Landau pole of the running of the dark sector strong interaction to this energy scale then allows us to make a projection of the impact of the running at large momentum transfers in the light of our discussion of Sec.~\ref{sec:results}. The results are given in Fig.~\ref{fig:spectra_landau}.

%%%%%%%%%%%%%%%%%%%%%%%%%%%%%%%%%
%%%%%%%%%%%%%%%%%%%%%%%%%%%%%%%%% 

As can be seen from Eq.~\eqref{eq:chixs}, if the self-interaction cross section is indeed dominated by the low-energy pion interactions, the cross section alone does not contain information about the strong dynamics as such (provided that the symmetry-breaking pattern indeed produces a spectrum that matches our assumptions). If this is the case, the only way to perform spectroscopy of the described scenario is through studies of the momentum dependence of the fundamental parameters of the dark sector UV theory. Since dark-gluon production is not directly accessible, an investigation through portal-interactions whose presence can be established through additional resonance searches is vital to gain information about the potential presence of such a sector given the discovery of an additional scalar which is compatible with a Higgs mixing scenario.

%\subsection{Achieving the correct relic density}

Although our main focus is the general behavior of general strongly interacting dark sectors and their spectroscopy using Higgs mixing, models with self-interacting hidden sectors should also reproduce the correct measured relic density $\Omega_{\textrm{DM}} h^2 \approx 0.12$ to be viable dark matter candidates. Our setup is flexible and allows for thermal freeze-out to occur either through standard annihilation into the SM, through number-changing $3 \to 2$ interactions between the dark pions as in~\cite{Hochberg:2014dra,Hochberg:2014kqa} (subject to the conditions detailed in this work), or a combination of the two, depending on the details of the chosen parameter point.

\medskip

There is also the possibility that glueballs make up most of the relic density instead of the pions as qualitatively discussed above. Since our discussion involves asymptotically free dark sectors, the analyses of~\cite{Boddy:2014yra,Soni:2016gzf} are applicable in this case: on the one hand, the correct relic density can be achieved by tuning the ratio of the visible and dark sector temperatures, which, however, requires an extremely small mixing. Under these circumstances the discovery of the additional scalar becomes impossible. On the other hand, if both sectors are in thermal contact through non-negligible mixing angles, we need to rely on additional (supersymmetric) dynamics to make the model cosmologically viable~\cite{Boddy:2014yra}. Our discussion does not apply in these cases straightforwardly and we leave an analysis of supersymmetric extensions to future work.\footnote{It is worthwhile mentioning that the authors of \cite{Boddy:2014yra} found that the number of colours is required to be small, which decreases the relative impact of the RGE running when the mixing angle interactions of hidden and visible sectors are non-neglible.}

%%%%%%%%%%%%%%%%%%%%%%%%%%%%%%%%%%%%%%%%%%%%%%%%%%
\section{Summary and Conclusions}
\label{sec:conc}
Dark sectors are SM extensions motivated to tackle a plethora of unexplained phenomenological observations that require physics beyond the SM. Their appeal from a model-building perspective comes at the price of a naturally suppressed phenomenological sensitivity yield in terrestrial experiments such as colliders. In this paper, using RGE-improved calculations, we have motivated that studying the energy dependence of scalar mediators, produced at a future hadron collider and decaying invisibly, can be utilised to gain some insights into the nature of the hidden sector, in particular because data-driven methods will be available for large data sets of $10~\text{ab}^{-1}$. Gaining excellent systematic control over the backgrounds well beyond the current expectations of theoretical as well as experimental uncertainties will be crucial to obtain these insights into strongly interacting dark sectors, which can complement other lattice investigations.

We have used this rather general observation for the concrete case of self-interacting dark matter, whose large cross section can be naturally explained by strong dynamics. If the strongly interacting dark matter scenario turns out to be true and its relation to the TeV scale through e.g. Higgs mixing becomes favoured, then the described approach will be a unique collider-based strategy that provides insight into a strongly interacting sector (supplied by calculations of finite corrections which are not governed in our RGE-based approach), albeit remaining experimentally challenging.

\bigskip 

\noindent{\bf{Acknowledgments}} 
--- We thank Joerg Jaeckel for helpful comments on the manuscript and Liam Moore for useful discussions and extensive 
help with \textsc{Mathematica}-related issues. MS would like to thank MIAPP ``Baryogenesis'' for hospitality during the finalisation of parts of this work. CE is supported in part by the IPPP Associate scheme. MS is supported in
part by the European Commission through the ``HiggsTools'' PITN-GA-2012-316704. KN is supported by the University of Glasgow College
of Science \& Engineering through a PhD scholarship.

%%%%%%%%%%%%%%%%%%%%%%%%%%%%%%%%%%%%%%%%%%%%%%%%%%
%%%%%%%%%%%%%%%%%%%%%%%%%%%%%%%%%%%%%%%%%%%%%%%%%%
%%%%%%%%%%%%%%%%%%%%%%%%%%%%%%%%%%%%%%%%%%%%%%%%%%
\bibliography{references}

\end{document}